\documentclass[fleqn,usenatbib]{mnras}

\usepackage[T1]{fontenc}
\usepackage{ae,aecompl}
\usepackage{times}
\usepackage{color}
\usepackage{soul}

\usepackage{graphicx}	
\usepackage{amsmath}	
\usepackage{amssymb}	

\newcommand{\cf}{cf.,~}
\newcommand{\ie}{i.e.,~}
\newcommand{\eg}{e.g.,~}

\title[Collapse to Kerr-Newman black hole]{Gravitational collapse to a
  Kerr-Newman black hole}

\author[Nathanail, Most, and Rezzolla]
{Antonios Nathanail$^{1}$ \thanks{E-mail: nathanail@th.physik.uni-frankfurt.de},
Elias R. Most$^{1}$,
Luciano Rezzolla$^{1,2}$
\\
$^{1}$Institut f\"ur Theoretische Physik, Goethe Universit\"at Frankfurt,
Max-von-Laue-Str.1, 60438 Frankfurt am Main, Germany\\
$^{2}$Frankfurt Institute for Advanced Studies, Ruth-Moufang-Str. 1, 60438
Frankfurt am Main, Germany 
}

\pubyear{2016}

\begin{document}
\label{firstpage}
\pagerange{\pageref{firstpage}--\pageref{lastpage}}
\maketitle

\begin{abstract}
We present the first systematic study of the gravitational collapse of
rotating and magnetised neutron stars to charged and rotating
(Kerr-Newman) black holes. In particular, we consider the collapse of
magnetised and rotating neutron stars assuming that no pair-creation
takes place and that the charge density in the magnetosphere is so low
that the stellar exterior can be described as an electrovacuum. Under
these assumptions, which are rather reasonable for a pulsar that has
crossed the ``death line'', we show that when the star is rotating, it
acquires a net initial electrical charge, which is then trapped inside
the apparent horizon of the newly formed back hole. We analyse a number
of different quantities to validate that the black hole produced is
indeed a Kerr-Newman one and show that, in the absence of rotation or
magnetic field, the end result of the collapse is a Schwarzschild or Kerr
black hole, respectively. 
\end{abstract}




\section{Introduction}
\label{sec:intro}

%
The rotating black-hole (Kerr) solution of the Einstein equations in
vacuum and axisymmetric spacetimes is a fundamental block in relativistic
astrophysics and has been studied in an enormously vast literature for
its mathematical and astrophysical properties. On the other hand, the
rotating and electrically charged black hole (Kerr-Newman, KN hereafter)
solution of the Einstein-Maxwell equations in axisymmetric spacetimes,
while equally well studied for its mathematical properties, is also
normally disregarded as astrophysically relevant. The rationale being
that if such an object was indeed created in an astrophysical scenario,
then the abundant free charges that accompany astrophysical plasmas would
neutralise it very rapidly, yielding therefore a standard Kerr solution.

Yet, KN black holes continue to be considered within astrophysical
scenarios to explain, for instance, potential electromagnetic
counterparts to merging stellar-mass binary black-hole systems
\citep{Zhang2016, Liebling2016, Liu2016}. We here take a different view
and do not explore the phenomenology of KN black holes when these are
taken to be long-lived astrophysical solutions. Rather, we are interested
to determine how such black holes are produced in the first place as, for
instance, in the collapse of rotating and magnetised stars. We note that
even if these solutions are short-lived astrophysically
\citep{Contopoulos2014, Punsly2016}, the study of their genesis can
provide useful information and shed light on some of the most puzzling
astronomical phenomena of the last decade: fast radio bursts [FRBs;
  \citet{Lorimer2007,Thornton2013}]. FRBs are bright, highly dispersed
millisecond radio single pulses that do not normally repeat and are not
associated with a known pulsar or gamma-ray burst. Their high dispersion
suggests they are produced by sources at cosmological distances and thus
with an extremely high radio luminosity, far larger than the power of
single pulses from a pulsar. The event rate is also estimated to be very
high and of a few percent that of supernovae explosions, making them very
common. Several theoretical models have been proposed over the last few
years, but the \textit{``blitzar''} model \citep{Falcke2013}, is
particularly relevant for our exploration of the formation of KN black
holes.

We recall that if a neutron star exceeds a certain limit in mass and
angular momentum, it will reach a state in which it cannot support itself
against gravitational collapse to a black hole. It is also widely
accepted that rotating magnetised neutron stars emitting pulsed radio
emission, \ie pulsars, spin down because of electromagnetic energy losses
and could therefore reach the stability line against collapse to a black
hole. During the collapse of such a pulsar, an apparent horizon is formed
which will cover all the stellar matter, while the magnetic-field lines
will snap violently launching an intense electromagnetic wave moving at
the speed of light. Free electrons will be accelerated by the travelling
magnetic shock, thus dissipating a significant fraction of the
magnetosphere energy into coherent electromagnetic emission and hence
produce a massive radio burst that could be observable out to
cosmological distances \citep{Falcke2013}.

One aspect of this scenario that has not yet been fully clarified is the
following: does the gravitational collapse of a rotating magnetised
neutron star lead to a KN black hole? The purpose of this paper is to
provide an answer to this question and to determine, through
numerical-relativity simulations, the conditions under which a collapsing
pulsar will lead to the formation of a Kerr or a Kerr-Newman black
hole. In particular, we show that when using self-consistent initial data
representing an unstable rotating and magnetised neutron star in general
relativity, the consequent collapse yields a black hole that has all the
features expected from a KN black hole. In particular, the spacetime
undergoes a transition from being magnetically dominated before the
collapse, to being electrically dominated after black-hole formation,
which is indeed a key feature of a KN black hole. We further provide
evidence by carefully analysing the Weyl scalar $\psi_2$ and by showing
that the black-hole spacetime possesses a net electric charge and a
behaviour which is the one expected for a KN black hole. These results
will be contrasted with those coming from the gravitational collapse of a
nonrotating magnetised neutron star, where the outcome is an uncharged
nonrotating (Schwarzschild) black hole.

The plan of the paper is the following one. In Sec. \ref{sec:nsaid} we
briefly review the numerical setup and how the initial data is computed,
leaving the analysis of the numerical results in
Sec. \ref{sec:nraa}. Finally, the discussion of the astrophysical impact
of the results and our conclusions are presented in Sec. \ref{sec:conc}

\section{Numerical setup and Initial data}
\label{sec:nsaid}

All simulations presented here have been performed employing the
general-relativistic resistive magnetohydrodynamics (MHD) code
\texttt{WhiskyRMHD} \citep{Dionysopoulou:2012pp,Dionysopoulou2015}, which
uses high-resolution shock capturing methods like the Harten-Lax-van
Leer-Einfeldt (HLLE) approximate Riemann solver coupled with effectively
second order piece-wise parabolic (PPM) reconstruction. Differently from
the implementation reported in Refs. \citep{Dionysopoulou:2012pp,
  Dionysopoulou2015}, we reconstruct our primitive variables at the cell
interfaces using the enhanced piecewise parabolic reconstruction (ePPM)
\citep{Colella2008, Reisswig2012b}, which does not reduce to first order
at local maxima. Also, we opt for reconstructing the quantity $Wv^i$,
where $W$ is the Lorentz factor, instead of the 3-velocity $v^i$; this
choice enforces subluminal velocities at the cell interface. Regarding
the electric-field evolution, we choose not to evolve the electrical
charge $q$ directly through an evolution equation, and instead compute
$q= \nabla_i E^i$ at every timestep, as it has been done by
\citet{Dionysopoulou:2012pp} and \citet{Bucciantini2012a}. The
\texttt{WhiskyRMHD} code exploits the Einstein Toolkit, with the
evolution of the spacetime obtained using the \texttt{McLachlan} code
\citep{loeffler_2011_et}, while the adaptive mesh refinement is provided
by \texttt{Carpet} \citep{Schnetter-etal-03b}.

The use of a resistive-MHD framework has the important advantage that it
allows us to model the exterior of the neutron star as an electrovacuum,
where the electrical conductivity is set to be negligibly small, so that
electromagnetic fields essentially evolve according to the Maxwell
equations in vacuum. These are the physical conditions that are expected
for a pulsar that has passed its ``death line'', that is, one for which
either the slow rotation or a comparatively weak magnetic field are such
that it is not possible to trigger pair creation and its magnetosphere
can be well approximated as an electrovacuum \citep{Chen1993}\footnote{We
  recall that the voltage drop $\Delta V$ along magnetic field lines
  needed for the creation of pairs scales with the magnetic field $B$ and
  rotation frequency $\Omega$ simply as $\Delta V \sim B\, \Omega$
  \citep{Ruderman1975}.}. At the same time, the resistive framework also
enables us to model the interior of the star as highly conducting fluid,
so that our equations recover the ideal-MHD limit (\ie infinite
conductivity) in such regions. \texttt{WhiskyRMHD} achieves this by
including a current that is valid both in the electrovacuum and in the
ideal-MHD limit, where it becomes stiff, however. To accurately treat
such a current, the code uses an implicit-explicit Runge-Kutta time
stepping (RKIMEX) \citep{pareschi_2005_ier}. For more details on the
numerical setup we refer the interested reader to
\citet{Dionysopoulou:2012pp} and \citet{Dionysopoulou2015}.

Our initial data is produced using the \texttt{Magstar} code of the
\texttt{LORENE} library, which can compute self-consistent solutions of
the Einstein-Maxwell equations relative to uniformly rotating stars with
either purely poloidal \citep{Bocquet1995} or toroidal magnetic fields
\citep{Frieben2012}; hereafter we will consider only poloidal magnetic
fields of dipolar type. We have considered a number of different possible
configurations for the electric field with the aim of minimising the
amount of external electric charges. In practice, the smallest external
charge has been achieved when prescribing a corotating interior electric
field matched to a divergence-free electric field produced by a rotating
magnetised sphere. More precisely, we set the electric field in the
stellar interior using the ideal-MHD condition, \ie
$E^i=-\sqrt{\gamma}\epsilon^{ijk} v_{j,{\rm cor}} B_k$ where $v_{j,{\rm
    cor}}$ is the corotation velocity, $\gamma$ the 3-metric determinant
and $\epsilon^{ijk}$ the totally antisymmetric permutation symbol. This
field is then matched to an exterior electrovacuum solution for a
magnetised and rotating uncharged sphere in general-relativity
\citep{Rezzolla2001, Rezzolla2001_err}. Note that because the analytic
solution is obtained in the slow-rotation approximation, which assumes a
spherical star, a small mismatch in the electric field is present near
the pole. Furthermore, monopolar and a quadrupolar terms are added to the
solution so as to match the charge produced by the corotating interior
electric field following \citet{Ruffini73}.

Also, as customary in this type of simulations, the stellar exterior is
filled with a very low-density fluid, or ``atmosphere'', whose velocity
is set to be zero \citep{Dionysopoulou:2012pp}; at the same time, and
from an electrodynamical point of view, we treat such a region as an
electrovacuum, so that the electrical conductivity is set to zero. This
has the important consequence that the magnetic fields are no longer
frozen in the atmosphere and are therefore free to rotate following the
stellar rotation if one is present.

Our reference rotating model is represented by a neutron star with
gravitational mass of $M=2.104\,M_{\odot}$, a period of $P=1.25\,{\rm
  ms}$ (or $800\,{\rm Hz}$), and a central (and maximum) magnetic field
of $10^{15}\,{\rm G}$; for such a model, the light cylinder is at about
$60\,{\rm km}$ from the origin.  The corresponding reference nonrotating
model has instead a gravitational mass of $M=2.100\,M_{\odot}$ and the
same magnetic field of $10^{15}\,{\rm G}$. Finally, we will also consider
a model with the same properties as the rotating one, but with zero
magnetic field. All models are constructed from a single polytrope $p=K
\rho^{\Gamma}$ with $\Gamma=2$. The polytropic constant $K=164.708$ has
been adjusted so that the maximum mass of a nonrotating star is limited
to about $2.1\,M_\odot$. The evolution is however performed using an
ideal-fluid equation of state $p=\rho\epsilon(\Gamma-1)$, where
$\epsilon$ is the specific internal energy. In spite of using a very
simplified equation of state, we do not expect this to have any effect on
the results of this paper since we are merely interested in a prompt
collapse to a black hole.

An important issue to discuss at this point is whether or not the star
possesses initially a net electrical charge. As it happens, the standard
solution provided by \texttt{Magstar} does have a net charge, although we
decided not to use such a solution as it is not the one leading to the
smallest external charge. At the same time, it is reasonable to expect
that the strong electromagnetic fields in a pulsar will not only generate
a charge separation, but they will also lead naturally to a net
charge. Assuming, that the rotating neutron star is endowed with a
dipolar magnetic field aligned with the rotation axis and that it is
surrounded by a ionised medium, it will induce a radial electric field
\citep{Cohen1975, Michel1999}.
\begin{equation}
E^r=B^\theta\frac{\Omega R\sin\theta}{c}
\approx B \frac{\Omega R}{c}\sin^2\theta\,, 
\label{Er}
\end{equation}
where $B$ is the equatorial value of the dipole magnetic field as
measured by a nonrotating observer, while $\Omega$ and $R$ are the
angular velocity and the radius of the star, respectively. As a result,
the net electric charge can be computed as
\begin{equation}
Q=\int_0^{\pi}2\pi R^2 \sin\theta E^r d\theta 
\approx\frac{8\pi}{3c}R^3 \Omega B\,.
\label{Q}
\end{equation}
In the stellar interior this charge is distributed so as to satisfy the
infinite-conductivity condition $\boldsymbol{E}\cdot \boldsymbol{B}=0$
everywhere. Stated differently, having a net charge is not necessarily
unrealistic, at least in this simplified model [see also
  \citet{Petri2012} and \cite{Petri2016}]; for our choice of initial
stellar model, Eq.  \eqref{Q} would yield $Q \approx 2.6 \times 10^{17}
C$.

We should also remark that the values we have chosen above for the
magnetic field and spin frequency are untypically high for a pulsar past
the death line. However, they are chosen to maximise the initial charge
in order to stabilise the numerical evolution and aid the final
determination of the charge from numerical noise. It is also simple to
check that a neutron star with such magnetic field and rapid rotation is
far from electrovacuum in its magnetosphere. However, we, believe that
this does not affect the general outcome of our simulations, which should
be viewed as a proof of concept. Finally, our numerical grid consists of
seven refinement levels extending to about $1075\,{\rm km}$, with a
finest resolution of $147\,{\rm m}$. Additional runs with resolutions of
$184, 220\,{\rm m}$, have been performed to test the consistency of the
results, but we here discuss only the results of the high-resolution
runs.

\section{Numerical results and Analysis}
\label{sec:nraa}

Overall, the gravitational collapse of our stellar models follows the
dynamics already discussed in detail by \citet{Dionysopoulou:2012pp} [see
  also \citet{Baumgarte02b2} and \citet{Lehner2011} for different but
  similar approaches], and the corresponding electromagnetic emission
under a variety of conditions will be presented by \citet{Most2017}. We
here focus our attention on comparing and contrasting the collapse of the
magnetised rotating and nonrotating models. Both stars are magnetised,
but only the rotating model possesses also an electric field induced by
the rotation; as a consequence of the presence/absence of this initial
electric field, the rotating/nonrotating star is initially
charged/uncharged.

Since the dynamics is rather similar in the two cases (the magnetic
fields and rotation speeds are still a small portion of the binding
energy), the differences between the two collapses is best tracked by the
electromagnetic energy invariant
\begin{equation}
\label{eq:B2_m_E2}
F_{\mu\nu}F^{\mu\nu}= 2(\boldsymbol{B}^2-\boldsymbol{E}^2)\,,
\end{equation}
where $F^{\mu\nu}$ is the Faraday tensor, while $\boldsymbol{B}$ and
$\boldsymbol{E}$ are respectively the magnetic and electric field
three-vectors measured by a normal observer. Being an invariant, the
quantity \eqref{eq:B2_m_E2} is coordinate independent and can provide a
sharp signature of the properties of the resulting black hole. We recall
that \eqref{eq:B2_m_E2} is identically zero for a Schwarzschild or a Kerr
black hole, while it is negative in the case of a KN black hole
\citep{Misner73}.

\begin{figure}
    \includegraphics[width=1.0\columnwidth]{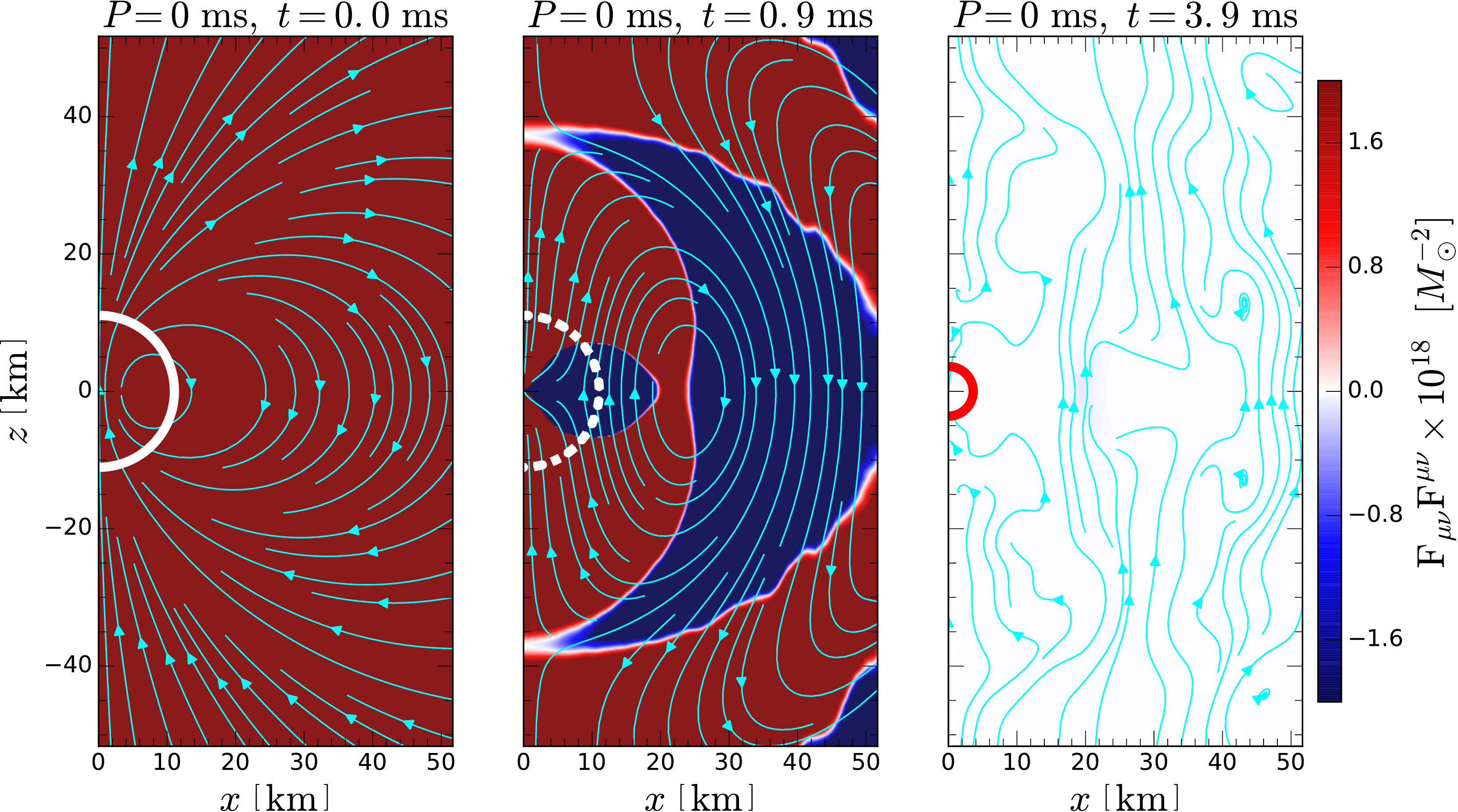}
    \includegraphics[width=1.0\columnwidth]{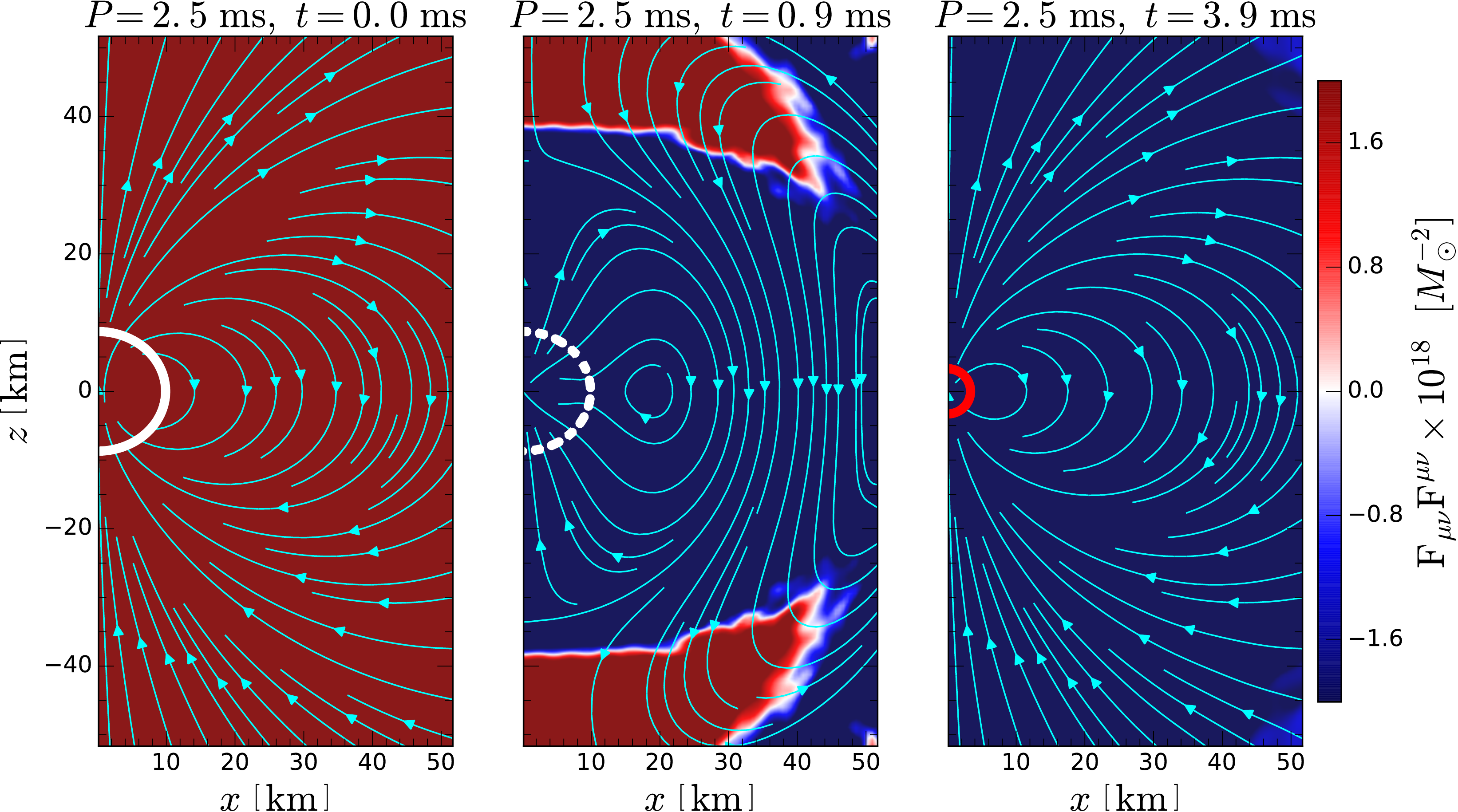}
     \caption{Evolution of the electromagnetic invariant
       $F_{\mu\nu}F^{\mu\nu}$; also shown are the stellar surface (solid
       and dashed white lines), the apparent horizon (solid red line),
       and the magnetic-field lines. The top and bottom rows refer to the
       nonrotating and rotating models, respectively; in either case the
       initial maximum magnetic field is $10^{15}\,{\rm G}$. Note that
       both models are magnetically dominated initially
       ($F_{\mu\nu}F^{\mu\nu}>0$) but that the nonrotating one yields a
       Schwarzschild black hole ($F_{\mu\nu}F^{\mu\nu}\simeq 0$), while
       the rotating model an electrically dominated KN black hole
       ($F_{\mu\nu}F^{\mu\nu}<0$).
       \label{fig:ULGRB2} }
\end{figure}

Figure \ref{fig:ULGRB2} summarises the dynamics and outcome of the
gravitational collapse by showing as colorcode the values of the energy
invariant $F_{\mu\nu}F^{\mu\nu}$ at three representative times (the
initial one, the final one and an intermediate stage). The top row, in
particular, refers to the nonrotating (but magnetised) stellar model,
while the bottom row shows the evolution in the case of the model
rotating at $800\,{\rm Hz}$; also shown are the magnetic field lines.

What is simple to recognise in Fig. \ref{fig:ULGRB2} is that the rotating
and nonrotating stars both start being magnetically dominated, \ie with
$F_{\mu\nu}F^{\mu\nu}> 0$ (top and bottom left panels). However, while
the collapse of the magnetised nonrotating star leads to a Schwarzschild
black hole for which $F_{\mu\nu}F^{\mu\nu} \simeq 0$ (top right panel),
the collapse of the magnetised rotating star yields an electrically
dominated black hole, \ie with $F_{\mu\nu}F^{\mu\nu} < 0$ (bottom right
panel). Furthermore, while electrically dominated regions are produced in
both collapses, these are radiated away in the case of a nonrotating star
[see also \citet{Dionysopoulou:2012pp}], in contrast with what happens
for the rotating star (\cf blue regions in Fig. \ref{fig:ULGRB2}).  Also
worth remarking in Fig. \ref{fig:ULGRB2} is that the magnetic field at
the end of the simulation becomes essentially uniform and extremely weak
[not shown, but see \citet{Most2017}] in the case of the nonrotating
model, while it asymptotes to a dipolar magnetic-field configuration in
the rotating case. Note that this field does not seem to have a neutral
point. This is consistent with the magnetic-field geometry of a KN black
hole \citep{Pekeris87}, where it can be imagined that the dipolar field
is generated by a ring-like current at the location of the ring
singularity of the corresponding Kerr black hole; our time and spatial
gauges prevent the appearance of such singularity and push it to the
origin of the coordinates.
\begin{figure}
  \begin{center}
    \includegraphics[width=1.0\columnwidth]{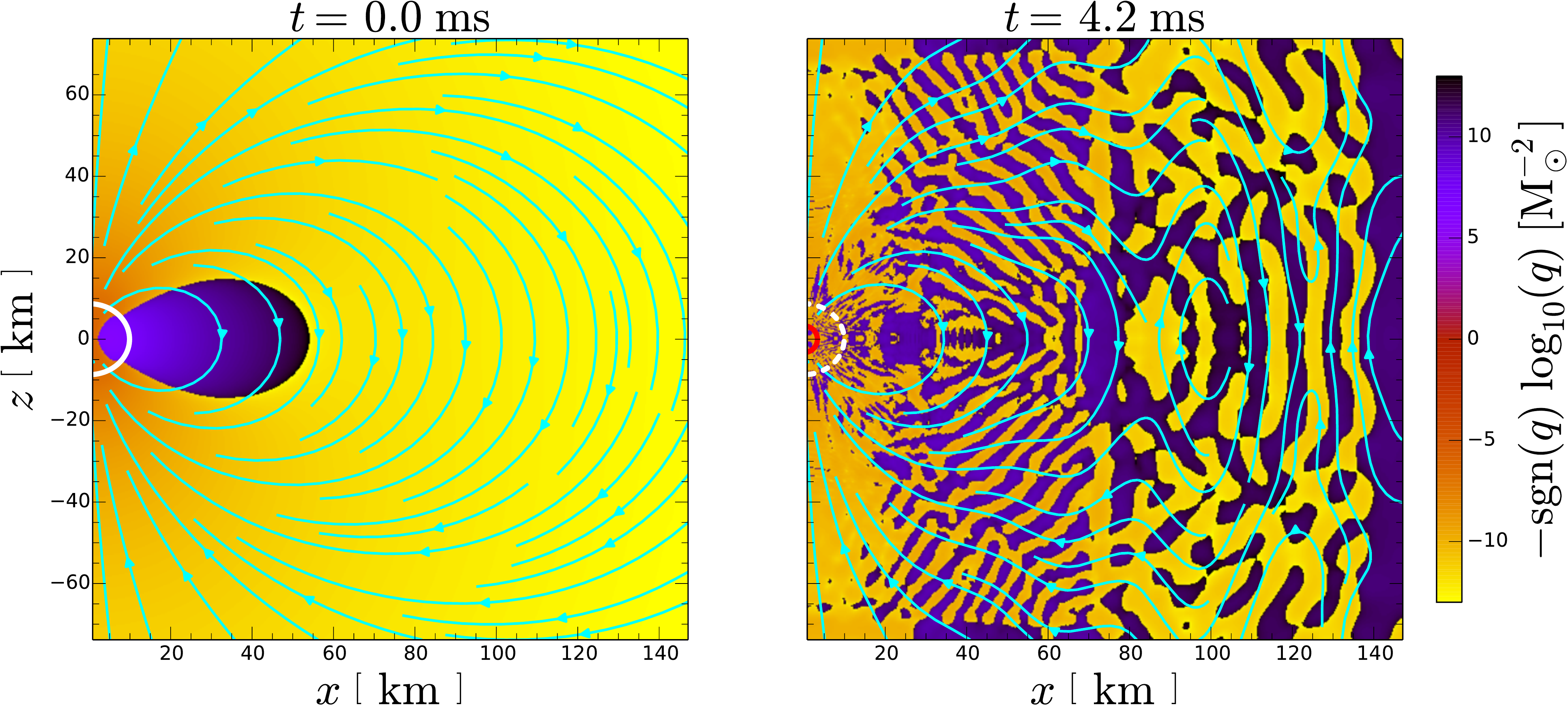}
     \caption{Charge density at the initial time (left panel) and after
       the collape (right panel); shown are also the stellar surface
       (solid/dashed white lines), the apparent horizon (red line), and
       the magnetic-field lines (cyan lines). Note the presence of a
       mesh-refinement boundary at $x\sim 70\,{\rm
         km}$.\label{fig:Qdens}}
  \end{center}
\end{figure}
As discussed in the previous section, the initial data contains a charge
density also in the stellar exterior, so that the overall charge in the
computational domain is given by the sum of the stellar charge and of the
exterior one; hereafter we refer to this charge as to $Q_{\rm
  tot}$. Shown in the two panels of Fig. \ref{fig:Qdens} is the
electrical charge distribution at the initial time (left panel) and at
the end of the simulation (right panel), together with the magnetic-field
lines, the location of the stellar surface (white solid and dashed lines
in the left and right panels, respectively) and of the apparent horizon
(red solid line in the right panel). Note that the initial charge density
falls off very rapidly with distance from the stellar surface and that
after a black hole has been formed, the charge is mostly dominated by
very small values with alternating signs; this behaviour is very similar
to the one observed when collapsing a nonrotating (uncharged) star and
hence indicates that the charge distribution in the right panel
is very close to the discretization error.

\begin{figure}
  \centering
  \includegraphics[width=0.98\columnwidth]{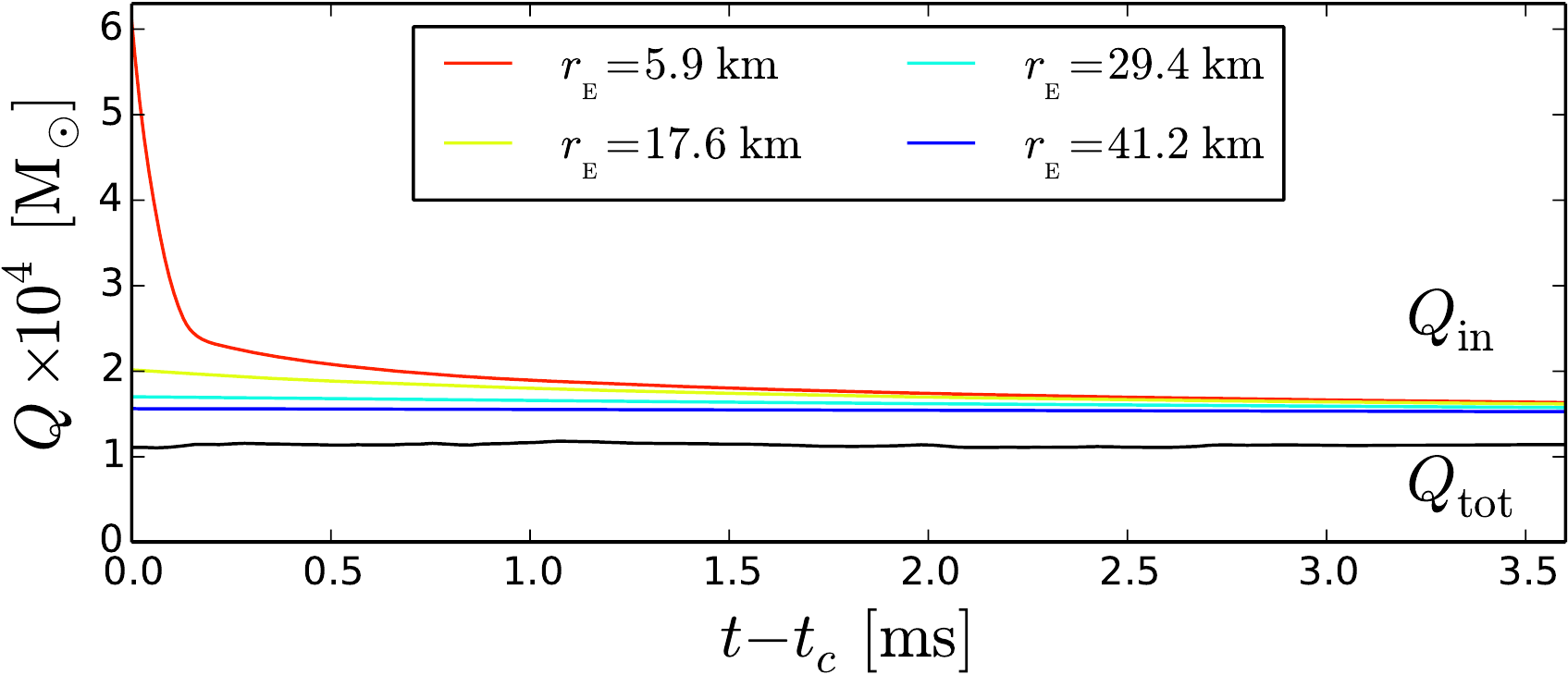}
  \caption{Evolution from the time of the formation of the apparent
    horizon, $t - t_c$, of the electric charge inside 2-spheres of radius
    $r_{_{\rm E}}$, $Q_{\rm in}$. Near the horizon the charge varies
    rapidly, reaching a constant value matched by larger 2-spheres. The
    black line refers to the the total charge in the domain, $Q_{\rm
      tot}$, which is well conserved. \label{fig:Qtot}}
\end{figure}

The presence of an external charge complicates the calculation of the
charge of the final black hole. In fact, when computing the total
electric charge as a surface integral of the normal electric field, we
inevitably include the contribution from the external charges. At the
same time, close to the horizon the calculation of the surface integral
is affected by numerical fluctuations, which deteriorate the accuracy of
the charge estimate. Hence, we measure the ``internal'' charge, $Q_{\rm
  in}$ by computing the integrals on successive 2-spheres of radius
$r_{_{\rm E}}$ and exploit the corresponding smooth behaviour to
extrapolate the value of the charge at the horizon. This is shown in
Fig. \ref{fig:Qtot}, where we report the evolution of the interior charge
$Q_{\rm in}$ as a function of the time from the formation of an apparent
horizon $t-t_c$, together with the total charge $Q_{\rm tot}$ computed
over the whole domain. Note that the latter is essentially constant over
time, thus indicating a good conservation of the total charge of the
system. Also, soon after the apparent horizon has formed a large portion
of the external charges, most of which are the result of the initial
mismatch in the electric field near the pole, is accreted onto the black
hole, leading to the rapid decrease of $Q_{\rm in}$ in
Fig. \ref{fig:Qtot} (red solid line).

Using the values of the interior charges from surfaces with a coordinate
radius $r_{_{\rm E}}=41.2\,{\rm km}$ down to $r_{_{\rm E}}=5.9\,{\rm
  km}$, we obtain an extrapolated value of the charge at the horizon
$Q_{_{\rm BH}}$ as the limit of $Q_{\rm in}(r_{_{\rm E}})$ for $r_{_{\rm
    E}} \to r_{_{\rm AH}}=3.26\,{\rm km}$. A simple quartic fit then
yields $Q_{_{\rm BH}} \simeq 1.6 \times 10^{-4}\, M_{\odot} \sim 1.87
\times 10^{16}\,{\rm C}$. This is a very large electrical charge, which
is the result of our initial data and probably not what should be
expected from a pulsar that has passed its death line. However,
determining such a charge under realistic conditions also requires an
accurate magnetospheric model, which still represents an open problem
despite the recent progress.

At this point it is not difficult to estimate the ``external'' charge
$Q_{\rm out}$ by subtracting the electrical charge trapped inside the
event horizon from the total one, \ie $Q_{\rm out}:=Q_{\rm tot}-Q_{\rm
  BH}\simeq -0.46 \times 10^{-4}\,M_{\odot} \sim -5.31 \times
10^{15}\,{\rm C}$. Note that $Q_{\rm out}$ is smaller than $Q_{\rm in}$,
but not much smaller and while $Q_{\rm in}$ is mostly positive, $Q_{\rm
  out}$ is mostly negative and present across the computational domain.
While the precise value we obtain for $Q_{_{\rm BH}}$ depends sensitively
on the initial electric field, the overall order of magnitude of the
charge is robust, as we discuss below. We can in fact validate that the
spacetime produced by the collapse of the rotating and magnetised star is
indeed a KN spacetime by considering a completely different
gauge-invariant quantity that is not directly related to electromagnetic
fields, but is instead a pure measure of curvature. More specifically,
for a KN black hole of mass $M_{_{\rm BH}}$, the only nonvanishing Weyl
scalar $\psi_2(r,\theta)$ is given by \citep{Adamo2014}
\begin{equation}
  \psi_2 = - \frac{M_{_{\rm BH}}}{\left(r -i a \cos
    \theta\right)^3} + \frac{Q^2_{_{\rm BH}}}{\left( r + i a \cos\theta
    \right)\left( r- i a \cos \theta \right)^3}\,. \label{eqn:psi2}
\end{equation}
This expression simplifies considerably on the equatorial plane (\ie for
$\theta=0$), where it becomes purely real and is 
\begin{equation}
  r^4 \psi_2 = -r\,M_{_{\rm BH}} + Q^2_{_{\rm BH}} \,.
  \label{eqn:psi2_equatorial}
\end{equation}
Because expression \eqref{eqn:psi2} holds true only in a pure KN
spacetime, which is not our case since our spacetime also contains a
small but nonnegligible external charge, we expect
\eqref{eqn:psi2_equatorial} to be more a consistency check than an
accurate measurement.

In practice, to distinguish the contribution in $\psi_2$ due to the mass
term from one due to the black-hole charge we compare the Weyl scalar
\eqref{eqn:psi2_equatorial} in two black holes produced respectively by a
rotating magnetised star and by a rotating non-magnetised star. Bearing
in mind that the magnetic field provides only a small contribution to the
energy budget, so that $\left. M_{_{\rm BH}} \right|_{B \neq 0} \simeq 
\left. M_{_{\rm BH}}\right|_{B=0}$ to a very good precision, we then obtain
\begin{align}
  Q^2_{_{\rm BH}} = r^4\left(\left. \psi_2 \right|_{B \neq 0} - 
  \left. \psi_2\right|_{B=0} \right)\,.
  \label{eq:QfromPsi2}
\end{align}
In principle, this quantity should be a constant in a pure KN solution,
despite $\psi_2$ being a function of position. In practice, in our
calculations this quantity has an oscillatory behavior around a constant
value in a region with $20 \lesssim r \lesssim 90\,{\rm km}$, while
higher deviations appear near the apparent horizon [where the spatial
  gauge conditions are very different from those considered by
  \citet{Adamo2014}] and at very large distances (where the imperfect
Sommerfeld boundary conditions spoil the solution locally). Averaging
around the constant value we read-off an estimate of the spacetime charge
from \eqref{eq:QfromPsi2} which is $Q_{_{\rm BH}} \approx 10^{-4}
M_{\odot}$. Given the uncertainties in the measurement, this estimate is
to be taken mostly as an order-of-magnitude validation of the charge of
the black hole $Q_{_{\rm BH}}$ measured as a surface integral. We expect
the precision of the geometrical measurement of the black-hole charge to
improve when increasing the resolution for the outer regions of the
computational domain and considerig longer evolutions that would lead to
a more stationary solution. In summary, by using a rather different
measurement based on curvature rather than on electromagnetic fields, we
converge on the conclusion that the collapse of the rotating magnetised
star leads to a KN black hole with a charge that is of a few parts in
$10^{4}$ of its mass for the initial data considered here.

\section{Conclusion}
\label{sec:conc}

We have carried out a systematic analysis of the gravitational collapse
of rotating and nonrotating magnetised neutron stars as a way to model
the fate of pulsars that have passed their death line but that are too
massive to be in stable equilibrium. The initial magnetised models are
the self-consistent solution of the Einstein-Maxwell equations and when a
rotation is present, they possess a magnetosphere and an
initial electrical net charge, as expected in the case of ordinary
pulsars. By using a resistive-MHD framework we can model the exterior of
the neutron star as an electrovacuum, so that electromagnetic fields
essentially evolve according to the Maxwell equations in vacuum. This is
not a fully consistent description of the magnetosphere, but it has the
advantage of simplicity and we expect it to be reasonable if the charge
is sufficiently small as for a pulsar that has crossed the death line.

The gravitational collapse, which is smoothly triggered by a progressive
reduction of the pressure, will lead to a burst of electromagnetic
radiation as explored in a number of works \citep{Baumgarte02b2,
  Lehner2011, Dionysopoulou:2012pp} and could serve as the basic
mechanism to explain the phenomenology of fast radio bursts
\citep{Falcke2013}. The end product of the collapse is either a
Schwarzschild black hole, if no rotation is present, or a KN black hole
if the star is initially rotating. For this latter case, we have provided
multiple evidence that the solution found is of KN type by considering
either electromagnetic and curvature invariants, or by measuring the
charge contained inside the apparent horizon. Hence, we conclude that the
production of a KN black hole from the collapse of a rotating and
magnetised neutron star is a robust process unless the star has zero
initial charge.

At the same time, a number of caveats should be made about our approach.
Our simulations have a simplistic treatment of the stellar exterior and
no microphysical description is attempted. It is expected, however, that
a distribution of electrons and positrons could be produced during the
collapse through pair production, leading to a different evolution
\citep{Lyutikov:2011b}. These charges could reduce the charge of the
black hole and even discharge it completely, possibly leading to a radio
signal that could be associated with fast radio bursts \citep{Punsly2016,
  Liu2016}. Furthermore, in the case of a force-free magnetosphere filled
with charges, the outcome of the collapse will likely be different,
although still yielding a KN black hole. Also, if present magnetic
reconnection in the exterior could change the evolution of the
electromagnetic fields and have an impact on the evolution of the charge
density. As a final remark we note that the dynamical production of a KN
black hole should not be taken as evidence for the astrophysical
existence of such objects. We still hold the expectation that stray
charges will rapidly neutralise the black-hole charge, so that a KN
solution should only be regarded as an intermediate and temporary stage
between the collapse of a rotating and magnetised star, \eg a pulsar that
has crossed the death line, and the final Kerr solution.
\section*{Acknowledgements}
We thank the referee, I. Contopoulos, for his constructive criticism that
has improved the presentation and the content of this paper. It is a
pleasure to thank K. Dionysopoulou and B. Mundim for help with the
\texttt{WhiskyRMHD} code, and B. Ahmedov and O. Porth for useful
discussions.  Partial support comes from the ERC Synergy Grant
``BlackHoleCam'' (Grant 610058), from ``NewCompStar'', COST Action
MP1304, from the LOEWE-Program in HIC for FAIR, from the European Union's
Horizon 2020 Research and Innovation Programme (Grant 671698) (call
FETHPC-1-2014, project ExaHyPE). AN is supported by an Alexander von
Humboldt Fellowship. The simulations were performed on SuperMUC at
LRZ-Munich, on LOEWE at CSC-Frankfurt and on Hazelhen at HLRS in
Stuttgart.
\section*{}
\bibliographystyle{mnras}
\bibliography{smallref_}

\label{lastpage}
\end{document}